\documentclass[aps,prl,amsmath,twocolumn,superscriptaddress,showpacs]{revtex4-1}

\usepackage{graphicx}
\usepackage{color}
\usepackage[normalem]{ulem}

\newcommand{\Fref}[1]{Figure~\ref{#1}}
\newcommand{\fref}[1]{Fig.~\ref{#1}}
\newcommand{\etal}{{\it et al.}}
\newcommand{\ba}{BaFe$_2$As$_2$}

\newcommand{\spm}{$s_{\pm}$}
\newcommand{\spp}{$s_{++}$}
\newcommand{\tc}{$T_c$}
\newcommand{\ts}{$T_s$}

\newcommand{\ccm}{C/cm$^2$}

\begin{document}

\title{Enhancement of $T_c$ by point-like disorder and anisotropic gap in FeSe}

\author{S.~Teknowijoyo}
\affiliation{Ames Laboratory and Department of Physics $\&$ Astronomy, Iowa State University, Ames, IA 50011, USA}

\author{K.~Cho}
\affiliation{Ames Laboratory and Department of Physics $\&$ Astronomy, Iowa State University, Ames, IA 50011, USA}

\author{M.~A.~Tanatar}
\affiliation{Ames Laboratory and Department of Physics $\&$ Astronomy, Iowa State University, Ames, IA 50011, USA}

\author{J.~Gonzales}
\affiliation{Ames Laboratory and Department of Physics $\&$ Astronomy, Iowa State University, Ames, IA 50011, USA}

\author{A.~E.~B\"{o}hmer}
\affiliation{Ames Laboratory and Department of Physics $\&$ Astronomy, Iowa State University, Ames, IA 50011, USA}

\author{O.~Cavani}
\affiliation{Laboratoire des Solides Irradi\'s, \'{E}cole Polytechnique, CNRS, CEA, Université Paris-Saclay, 91128 Palaiseau Cedex, France}

\author{V.~ Mishra}
%\email{vivekm.phys@gmail.com}
\affiliation{Joint Institute of Computational Sciences, University of Tennessee, Knoxville, TN-37996, USA}
\affiliation{Center for Nanophase Materials Sciences, Oak Ridge National Laboratory, Oak Ridge, TN-37831, USA}

\author{P.~J.~Hirschfeld}
%\email{pjh@phys.ufl.edu}
\affiliation{Department of Physics, University of Florida, Gainesville, Florida 32611, USA}

\author{S.~L.~Bud'ko}
\affiliation{Ames Laboratory and Department of Physics $\&$ Astronomy, Iowa State University, Ames, IA 50011, USA}

\author{P.~C.~Canfield}
\affiliation{Ames Laboratory and Department of Physics $\&$ Astronomy, Iowa State University, Ames, IA 50011, USA}

\author{R.~Prozorov}
\email[Corresponding author: ]{prozorov@ameslab.gov}
\affiliation{Ames Laboratory and Department of Physics $\&$ Astronomy, Iowa State University, Ames, IA 50011, USA}

\date{13 May 2016}

\begin{abstract}
A highly anisotropic superconducting gap is found in single crystals of FeSe
by studying the London penetration depth, $\Delta \lambda$, measured down to 50 mK
in samples before and after 2.5 MeV electron irradiation. The gap minimum increases
with introduced point - like disorder, indicating the absence of
symmetry - imposed nodes. Surprisingly, the superconducting transition temperature,
$T_c$, \textit{increases} by 0.4 K from $T_{c0} \approx$ 8.8 K while the structural transition
temperature, $T_s$, \textit{decreases} by 0.9 K from $T_{s0} \approx$91.2 K after electron irradiation.
We discuss several explanations  for the $T_c$ enhancement, and propose that local strengthening of the
pair interaction by irradiation-induced Frenkel defects most likely explains the phenomenon.
\end{abstract}

\maketitle

\section{Introduction}
Deliberately introduced point - like disorder may serve as a \emph{phase - sensitive} tool
to probe the superconducting gap structure and relative amplitudes of the pairing potential \cite{Hirschfeld1993,Prozorov2006SST,EfremovPRB2011,TcEnhancement2012,WangHirschfeldMishra2013PRB,Prozorov2014PRX}.
Usually, only the changes of the superconducting transition temperature, $T_c$,
are studied. However, in complex materials, such as iron-based superconductors
(IBS), this does not lead to unique predictions, see Ref.~\onlinecite{Prozorov2014PRX} and
references therein. Therefore, simultaneous measurement of another disorder-sensitive parameter, for example, London penetration depth, $\lambda(T)$,
can be used to impose additional constraints on the possible pairing models.
Measurements of the low-temperature variation, $\Delta \lambda(T) = \lambda(T) - \lambda(0)$,
can be used to study the gap anisotropy \cite{Strehlow2014,Cho2014PRB} and to distinguish between \spm\ and $s_{++}$ pairing \cite{WangHirschfeldMishra2013PRB}.
The latter was successfully used to study nodal BaFe$_2$(As,P)$_2$ \cite{Mizukami2014NatureComm}
and SrFe$_2$(As,P)$_2$ \cite{Strehlow2014} where potential scattering lifted the
nodes proving them accidental, therefore strongly supporting \spm\ pairing.

The majority of iron - based superconductors (IBS) have a region of coexisting
superconductivity and long-range magnetic order (LRMO) in their temperature -
composition phase diagram, usually at low doping levels. Whereas this leads to some
very interesting physics \cite{Canfield2010review122,Mazin2010,Johnston2010review,Paglione2010review,ProzorovKogan2011RPP,StewartRMP2011,ChubukovAR2012,ChubukovHirschfeld2015,Hirschfeld2016}, it complicates the analysis of the superconducting gap structure \cite{Fernandes2010,TcEnhancement2012,ProzorovKogan2011RPP}.
FeSe, on the other hand, only exhibits a structural transition around
$T_s \approx$ 90 K, but no LRMO at the ambient pressure \cite{Mizuguchi2010}. Being
a nearly stoichiometric compound with relatively simple electronic band structure \cite{Subedi2008},
FeSe offers a unique opportunity to study iron-based superconductivity without complications
of LRMO and elevated scattering, which is always significant in charge - doped
compounds \cite{Kogan2009,Gordon2010a,ProzorovKogan2011RPP}. The temperature-pressure phase
diagram of FeSe is quite non-trivial. The superconducting transition
temperature, $T_c$, is non-monotonic, -- increasing initially up to 0.8 GPa, then decreasing,
reaching  a minimum at 1.3 GPa and increasing again \cite{Miyoshi2014,KnonerPRB2015,Kaluarachchi2016}.
Despite the absence of LRMO, a strong nematic response is found in FeSe, and
has been discussed in terms of both spin and orbital fluctuations\cite{Kivelson2015,Glasbrenner2015,Kontani2015,Chubukov2016,Tanatar2016}.
Additional interest in this material stems from the discovery of
high temperature superconductivity with $T_c \approx$ 65 K in a single-layer FeSe
grown on a SrTiO$_3$ \cite{Liu2012,He2013}, as well as the intriguing possibility of
being in the regime of a crossover from Bose-Einstein condensation (BEC) and
Bardeen-Cooper-Schrieffer (BCS) coupling due to small Fermi energies comparable to superconducting gap values \cite{Kasahara2014,Terashima2014}.

Most  studies of IBS have converged on generalized \spm\ pairing as the basic and quite robust pairing mechanism supporting both nodeless
and nodal states \cite{ChubukovHirschfeld2015,Hirschfeld2016}. In FeSe, anisotropic line nodes or deep minima were found theoretically \cite{Choubey2014,Kreisel2015,Mukherjee2015}.
Experimentally, scanning tunneling spectroscopy (STS) \cite{Song2011},
London penetration depth and thermal conductivity \cite{Kasahara2014} claimed
nodal superconductivity. However, measurements of the lower critical field \cite{Hc1FeSe2013},
low-temperature specific heat \cite{Lin2011,Jiao2016}, other STM \cite{Jiao2016} and other thermal
conductivity studies \cite{Dong2009,Bourgeois2016} are consistent with nodeless
superconducting gap. A cross-over from nodal in the bulk to nodeless at the
twin boundary is found from STS \cite{Watashige2015}. In all these studies,
however, highly anisotropic gap and/or multiband physics are present.
On the other hand, a single large
nodeless gap has been reported in single-layer FeSe \cite{Liu2012,LiuJPCM2015}. Despite the same chemical formula, this material also has a very different bandstructure and very different \tc, compared to the bulk FeSe. This, however, shows how susceptible this compound is to permutations of its chemical-physical state.

\begin{figure}[htb]
\centering
\includegraphics[width=8.5cm]{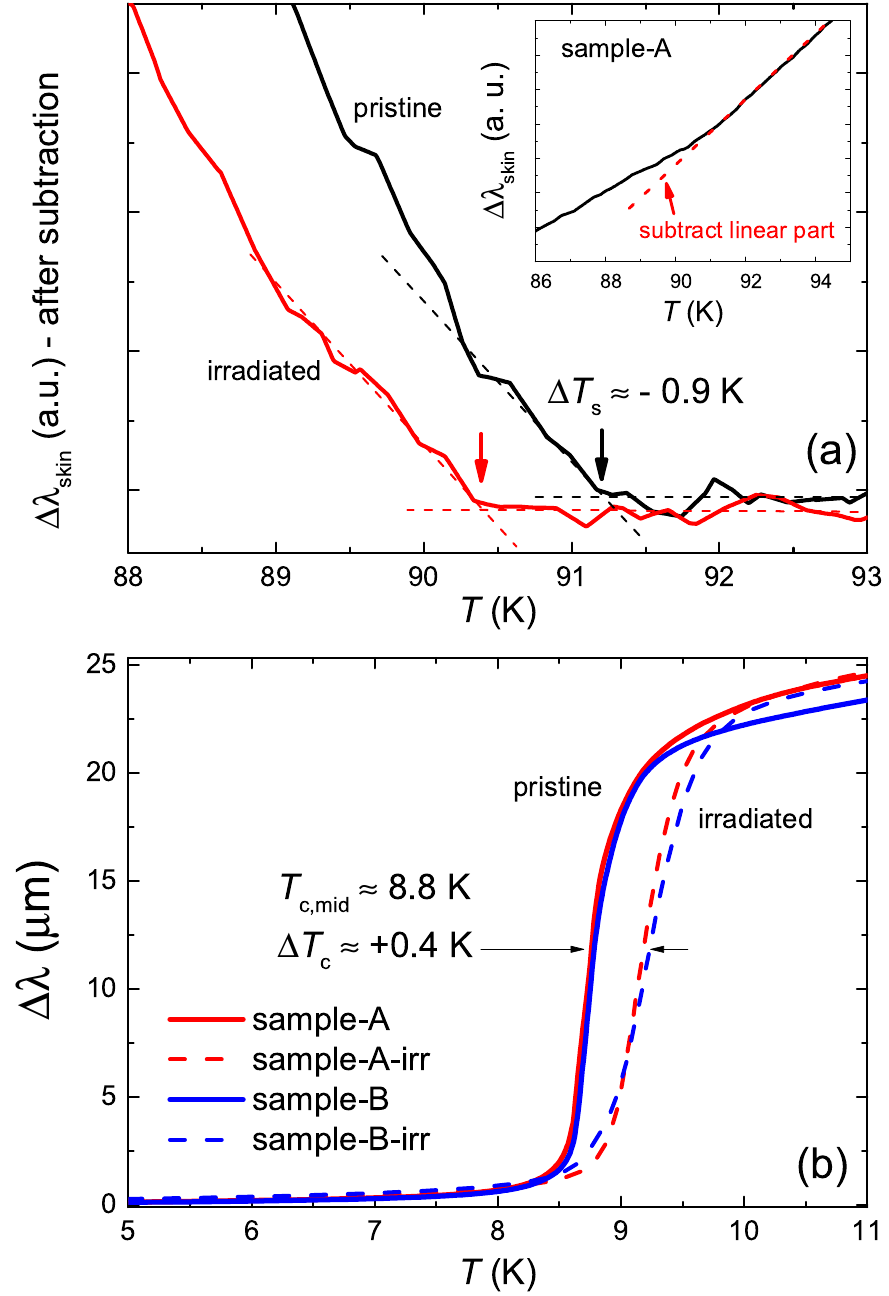}
\caption{(Color online) (a) Variation of normal state skin depth $\Delta \lambda_{skin} (T \gg T_c)$
of sample A  after linear-subtraction as shown in the \emph{inset}. Arrows mark
structural transition, $T_s$, before and after 2.5 MeV electron irradiation of
1.8 C/cm$^2$. (b) London penetration depth $\Delta \lambda (T)$ before and after
electron irradiation in Samples A and B.}
\label{fig1}
\end{figure}

One possible scenario to reconcile these apparently contradictory results is
to consider marginal, accidental nodes in the clean limit, which are lifted by
the natural disorder always present to some degree in actual samples \cite{Bourgeois2016}.
To probe this scenario, in this work the superconducting gap structure of vapor-transport
grown FeSe crystals was studied by measuring the low-temperature variation of the London
penetration depth, $\Delta \lambda (T)$, before and after 2.5 MeV electron irradiation.
Using power - law fitting, $\Delta \lambda (t) \sim t^n$, ($t \equiv T/T_c$), we find
that the exponent $n$ is much greater than the terminal dirty-limit value of 2 in all
samples, signaling a nodeless gap. Irradiated samples show an even larger $n$, extending
up to a higher temperature of the fitting range. Moreover, BCS - like fitting with gap magnitude as free parameter clearly
shows an increase of the gap minimum upon introduction of point - like disorder. Both
results are consistent with the smearing of the anisotropic part of the gap. Surprisingly,
after $1.1 \times 10^{19}$ $e^-$/cm$^2$ 2.5 MeV electron irradiation,
$T_c$ has \emph{increased} by 0.4 K from 8.8 K, while $T_s$ decreased by 0.9 K from 91.2 K.
These opposite trends are similar to the effect of pressure \cite{KnonerPRB2015,Kaluarachchi2016,Dongna2016} and also imply that pair-breaking due to non-magnetic disorder is quite small.
Overall, our results are consistent with highly anisotropic superconducting gap, which may have accidental nodes in the clean limit. While
we cannot distinguish between generalized \spm\ and highly anisotropic multiband $s_{++}$ pairing, we can limit the former to the case where
intra-band pairing dominates the inter-band pairing.

Throughout the paper, we use the following terminology for multiband pairing: \spp\ is when the superconducting order parameters are of the
same sign on different bands, and \spm\ when some are of the opposite sign.
For the latter we distinguish between the case of dominant intraband pairing
vs. dominant interband pairing, since these two cases respond very
differently to nonmagnetic disorder.   For a 2 - band system with interaction
potential $V_{ij}$, the former is realized when $<V> \equiv n_1(V_{11}+V_{12})+n_2(V_{22}+V_{21})>0$,
where $V_{11}$ and $V_{22}$ are intraband, and $V_{12}$ and $V_{21}$ are inter-band
pairing potentials and $n_1=N_1/N(0)$ and $n_2=1-n_1$ are the normalized partial
densities of state (DOS) on two bands and $N(0)$ is the total DOS \cite{EfremovPRB2011}.
We will call this state ``intraband" \spm. The second possibility, $<V> < 0$,
is ``interband" \spm. It is important to note that even when $<V> > 0$, the order
parameters will have opposite signs and, thus, this is an \spm\ pairing state.

Finally, we note that the term s-wave pairing used throughout this
paper refers to the state that has
the full symmetry of the lattice just above the superconducting transition.
In the case of FeSe, this is $C_2$ rather than $C_4$ due to the strong
nematic symmetry breaking that occurs at the structural transition, and
the Fermi surface that drives the superconducting gap function is strongly
$C_2$ symmetric according to ARPES.  In
terms of the harmonics of the tetragonal system, such a state would be described as an $s+d$ state

\section{Experimental}
Single crystals of FeSe were grown using a modified chemical vapor transport method \cite{Bohmer2013,Kaluarachchi2016}. The variation of the
in-plane London penetration depth, $\Delta \lambda (T)$, was measured using a self-oscillating tunnel-diode resonator (TDR) down to 50 mK \cite{Prozorov2000PRB,Prozorov2006SST, ProzorovKogan2011RPP}.
The crystals under study have typical dimensions of about 0.5 $\times$ 0.5 $\times$ 0.03 mm$^3$. The samples were extensively characterized by measurements of magnetization, electric transport, M\"{o}ssbauer spectroscopy and high energy x-ray scattering, including under pressure as described elsewhere \cite{Kaluarachchi2016,Tanatar2016,Kothapalli2016Tu}.

The ratio of resistivities, $RRR$(300/10)$\equiv \rho$(300 K)$/\rho$(10 K)$\approx 20$. A simple linear extrapolation
to $T=0$, gives $RRR$(300/0)$\approx$ 125. In comparison, previous work on vapor transport grown samples
that found nodal superconductivity gives a very similar for $RRR$(300/10), but results in a negative linear
extrapolation, indicating lower residual resistivity, $\rho (0)$, hence a potentially less disordered sample \cite{Kasahara2014}.

To investigate the effect of deliberately introduced point - like disorder, $\Delta \lambda (T)$ was measured before and after 2.5 MeV electron irradiation performed at the SIRIUS Pelletron facility of the Laboratoire des Solides Irradies (LSI) at the \'{E}cole Polytechnique, France \cite{SIRIUS}. The acquired irradiation dose for our two irradiated samples was 1.8 C/cm$^2$. Here 1 C/cm$^2 =$ 6.24 $\times 10^{18}$ electrons/cm$^2$.
As shown in the \emph{Supplementary Information} section, by calculating the Frenkel pairs
(vacancy - interstitial) production cross-section we estimate creation of $\sim$0.05 at.\% of Frenkel
pairs per Fe and per Se (0.1 at.\% total pairs per formula or 0.2 at.\% per unit cell (Z=2)). Within the excellent sensitivity of the TDR technique,
these defects are non-magnetic.

Three samples were measured. Samples A and B were measured before and after
electron irradiation. Sample C was measured, cut in half and measured again to
estimate the $c-$axis London penetration depth as described in Ref.~\onlinecite{ProzorovKogan2011RPP}.

\section{Results}

\Fref{fig1} shows high temperature measurements to probe the
effect of electron irradiation on $T_c$ and $T_s$. In the normal state,
the TDR signal is proportional to the normal skin depth, $\lambda_{skin} \sim \sqrt{\rho}$ and the
resistivity, $\rho (T)$, has a kink at $T_s$ \cite{Tanatar2016} which is detected here
via $\lambda_{skin}(T)$. To visualize the transition, we subtract a linear part above $T_s$ as
shown in the \emph{inset} in \fref{fig1}(a). The structural transition temperature, $T_s$,
has shifted down by -0.9 K in sample A after irradiation. Similar behavior was also observed
for sample B. \Fref{fig1}(b) shows the region of superconducting transition. Both samples A
and B show very similar behavior with $T_c \approx $8.8 K (mid-point) \emph{increasing} by 0.4 K. Such
increase is highly unusual and its observation imposes strict limitations on the structure
of the superconducting order parameter. We note that although \tc\ enhancement reported here was measured in two different samples, we only had the opportunity to access one irradiation dose of 1.8 \ccm. Scenarios discussed in this paper may, in fact, lead to some non-monotonic behavior and further studies of \ts\ and \tc\ as function of irradiation dose are needed.

\begin{figure}[htb]
\centering
\includegraphics[width=8cm]{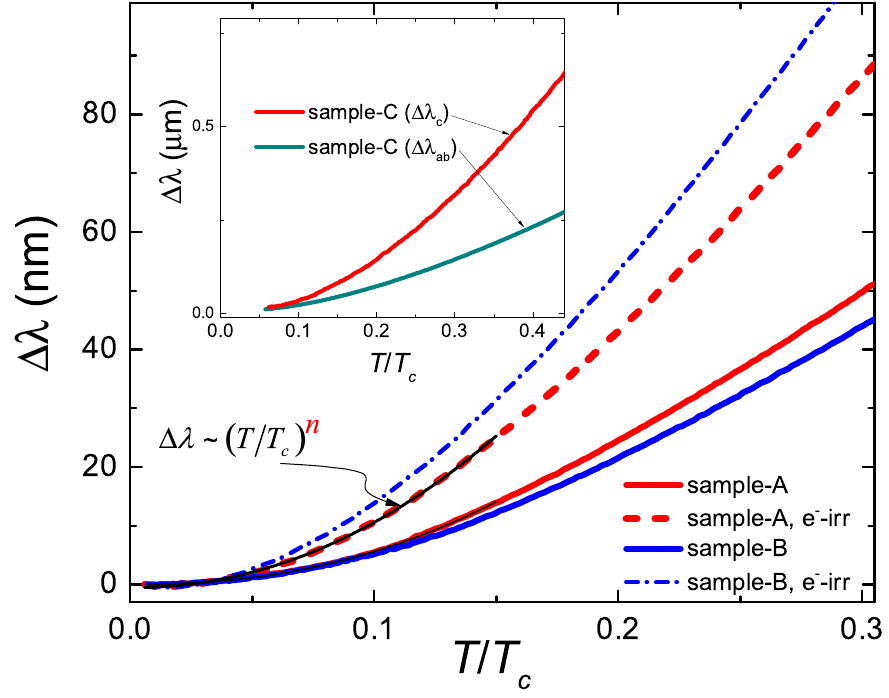}
\caption{(Color online) Low-temperature part of $\Delta \lambda (t)$ of samples
A (red) and B (blue) before (solid lines) and after (dashed lines) electron
irradiation of 1.8 C/cm$^2$. \emph{Inset} shows $\Delta \lambda _{ab}$ (teal) and $\Delta \lambda _{c}$ (red) of sample C.}
\label{fig2}
\end{figure}

\Fref{fig2} shows $\Delta \lambda (t)$ of samples A and B before (solid lines)
and after (dashed lines) 2.5 MeV electron irradiation dose of 1.8 \ccm. The penetration depth
remains practically flat at $T < 0.05 T_c$. Its amplitude increases faster with temperatures
in irradiated samples, signaling an increase of the number of thermally excited quasiparticles
compared to the pristine case. The inset shows in-plane, ($\Delta \lambda _{ab}$), and out-of-plane,
($\Delta \lambda _{c}$), penetration depths measured in sample C \cite{ProzorovKogan2011RPP}.
The ratio of $\Delta \lambda _{ab}$ and $\Delta \lambda _{c}$ at $T = 0.3 T_c$ is about 3,
consistent with the relatively low anisotropy of other iron-based superconductors \cite{ProzorovKogan2011RPP}.

\begin{figure}[htb]
\includegraphics[width=8cm]{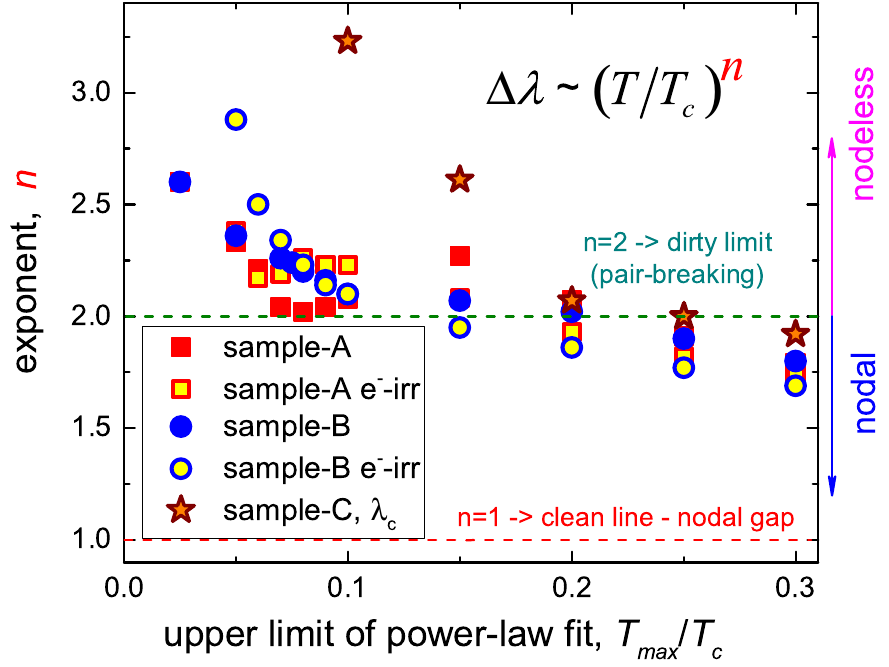}
\caption{(Color online) Exponent, $n$, of power-law fitting for data shown
in \fref{fig2}. x-axis is the upper-limit of the fitting range. In
all samples, the exponents increase well above dirty-limit of $n = $2 at
low-temperatures, indicating the presence of small, but finite superconducting
gap. After electron irradiation, $n$ becomes even higher, probably signaling
some reduction of the gap anisotropy. The $c-$ axis direction is also gapped.}
\label{fig3}
\end{figure}

With an apparent saturation of $\Delta \lambda (T)$ only at quite low temperatures, we
analyze its behavior using two approaches. First, following our previous studies \cite{ProzorovKogan2011RPP},
we fit the London penetration depth by the power-law, $\Delta \lambda (t) = A t^n$. The solid black curve,
indicated by an arrow in \fref{fig2}, shows an example of such a fit. We examine the dependence
of the exponent $n$ on the upper limit of the fitting range, $T_{max}/T_c$, which was varied
from 0.05 $T_c$ to 0.3 $T_c$ while the lower-limit was fixed as a base temperature of about 50 mK.
\Fref{fig3} shows how the exponent $n$ increases with the decrease of $T_{max}/T_c$
reaching the values significantly greater than 2 below 0.1$T_c$. This indicates the
presence of a small but finite gap, because both accidental and symmetry-imposed
line nodes result in $1 \leq n \leq 2$. %

As discussed above, STS experiments on high quality samples reported evidence
for gap nodes in thin films \cite{Song2011} and single crystals \cite{Kasahara2014}, and from the theoretical standpoint,
a ground  state with very shallow $C_2$-symmetric nodes was found within spin fluctuation calculations with orbital
ordering\cite{Mukherjee2015,Kreisel2015}, both in apparent contrast to our small
gap result \footnote{Note that because our samples are twinned \cite{Tanatar2016},
the supercurrent flows through structural domains of both orientations. In principle,
if one orientation is nodal, we should detect it in the clean limit.}.  However, we
know that  accidental nodes can be lifted by intraband disorder scattering\cite{Mishra2009PRB}.
It may therefore be that our samples are slightly more disordered than
those that show nodes. A similar suggestion was made in recent work on thermal conductivity \cite{Bourgeois2016}.

It is also possible that samples of FeSe differ from one  another not because of small differences
in defect concentrations, but due to different concentrations of twin boundaries
due to growth conditions.  Watashige \etal~\cite{Watashige2015} have shown that even
the bulk  crystals exhibiting a nodal state show large scale regions of full gap behavior
in the neighborhood of twin boundaries.  Depending on its irregularity, the twin boundary
may act a pair breaker, in which case this effect may be simply another version of the
disorder node lifting phenomenon. The long range nature of the effect\cite{Watashige2015}
suggests, however, that other physics may be in play.  At present we cannot make convincing
statements about the origin of our small gaps, but it appears clear that the gap is sensitive
to small perturbations, which can gap a nodal state, and at present the most natural
explanation seems to be that   disorder  is lifting the nodes in slightly less pure samples.

\begin{figure}[htb]
\includegraphics[width=8cm]{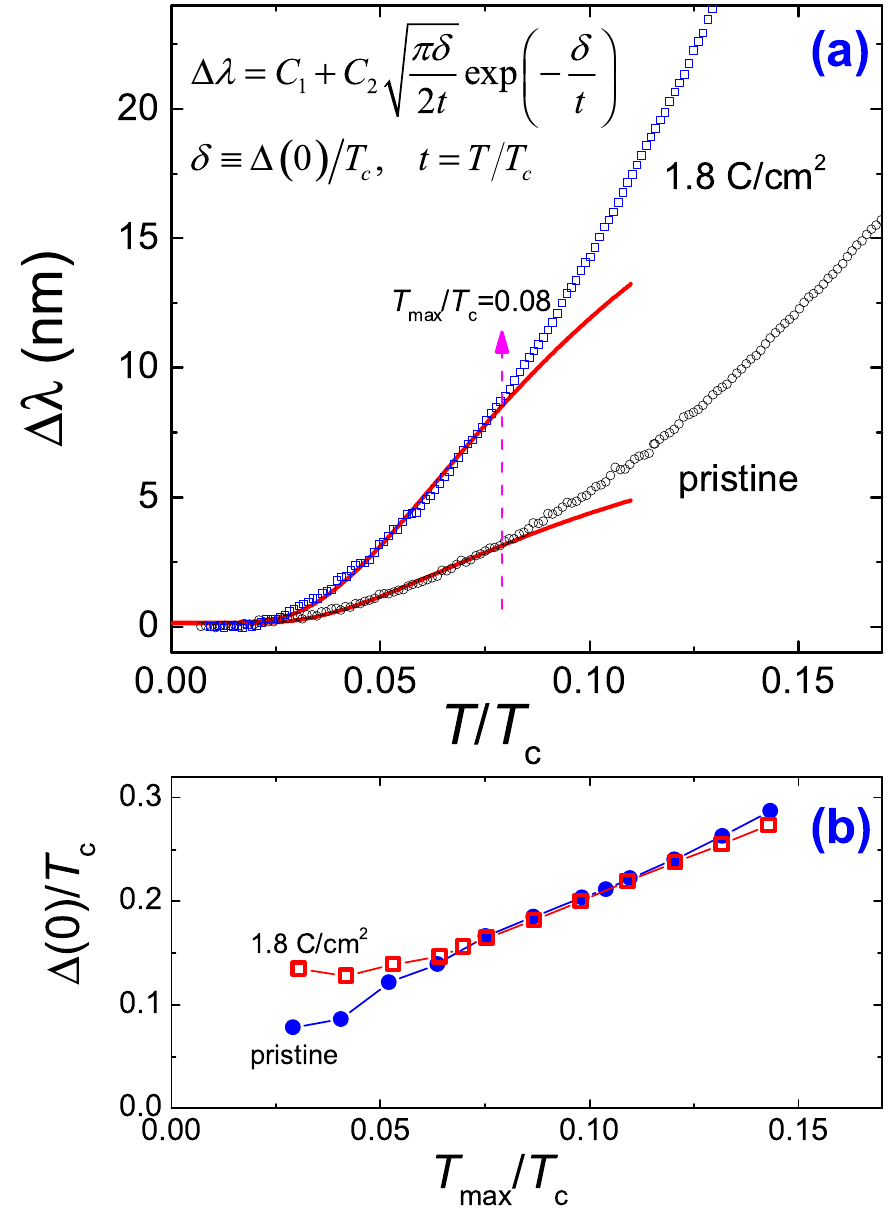}
\caption{(Color online) (a) low temperature $\Delta \lambda (t)$ and example of BCS - like fitting
of data for sample B for $T_{max}/T_c \approx $0.08 before (lower curve) and after 1.8 \ccm\ electron
irradiation (upper curve). Also shown an equation and definitions used. (b) $\Delta (0)/T_c$ ratio
obtained as a best fit parameter with different upper limits of the fitting range.}
\label{fig4}
\end{figure}

Our second approach to analyze low-temperature behavior is to
use  BCS single gap fit, $\Delta \lambda = C_1 + C_2 \sqrt{\pi \delta/2t}\exp{(-\delta/t)}$
with variable upper temperature limit, $T_{max}/T_c$, free parameters $C_1$ and $C_2$,
and the value of the gap, $\delta=\Delta(0)/T_c$ also as a free parameter.
This procedure can be used to estimate the minimum gap in the system, provided
that the measurements were done down to low enough temperature, which is the case
here. \Fref{fig4}(a) shows one example of the exponential fitting of the sample B
data before and after electron irradiation. \Fref{fig4}(b) presents the ratio
of $\Delta (0)/T_c$ obtained as the best fit parameter for several values of
the upper limit of the fitting range. While there is only a hint of
saturation in the pristine curve, the irradiated fits saturate at
about $\Delta_{min} (0)/T_c$ indicating a truly exponential behavior.
In addition, we see that the smaller range fits indicate clearly
that the minimum gap has increased upon irradiation, a phenomenon
analogous to node lifting, which results from the averaging
of the gap anisotropy by intraband disorder\cite{Mishra2009PRB}.
This is only possible if the anisotropy, and possible nodes, are not
imposed by the pairing potential symmetry (i.e., anisotropic $s-$ wave or \spm, but not  $d-$wave).

\begin{figure}[htb]
	\includegraphics[width=8.5cm]{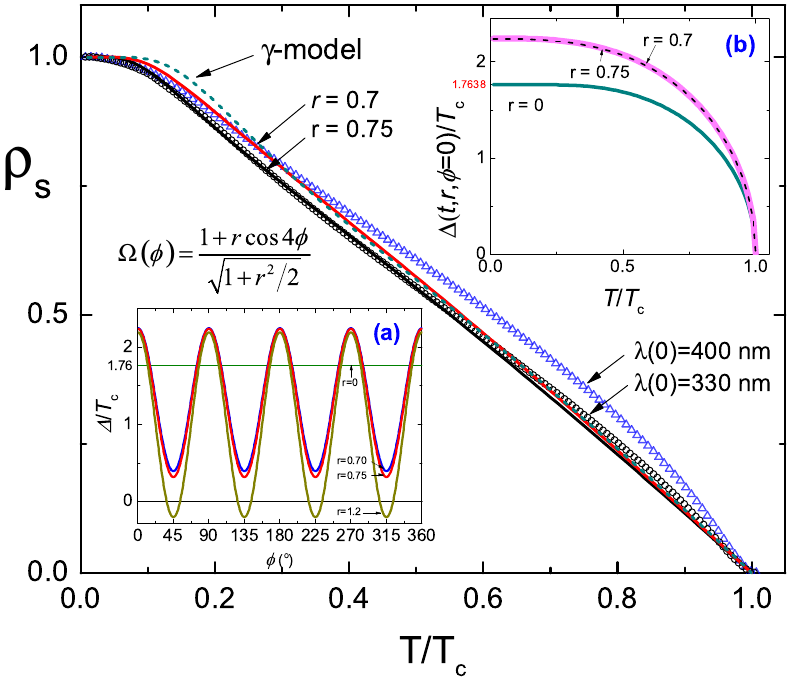}
	\caption{(Color online) Superfluid density analyzed in terms
	of anisotropic gap in the form shown. The best fit is obtained
	for $\rho_s$ calculated with $\lambda(0)=$ 330 nm, but $\rho_s$ with $\lambda(0)=$400 nm
	can also be fitted at $t < 0.3$. For completeness, a two-gap $\gamma-$model\cite{gammaModel}
	fit is shown by the dashed line. \emph{Inset} (a) angular variation of the gap with $r=$0.70
	(best fit of $\lambda(0)=$400 nm data), $r=$0.75 (best fit of $\lambda(0)=$ 330 nm data)
	and $r=$1.2 of the hypothetical accidental nodes state. \emph{Inset} (b) shows variation of
	the gap with temperature, obtained from the self-consistency equation.}
\label{fig5}
\end{figure}

To gain further insight into the gap structure, we need to analyze
the temperature-dependent superfluid density, $\rho_s=\left(1+\Delta \lambda (T)/\lambda(0)\right)^{-2}$,
over the entire temperature interval. Our TDR technique only measured $\Delta \lambda(T)$
and we need to know the absolute value of the London penetration depth, $\lambda(0)$.
In \fref{fig5}, the superfluid density $\rho_s (t)$ is plotted with $\lambda (0) =$ 400 nm
obtained from microwave cavity perturbation measurements of similar FeSe crystals \cite{Kasahara2014}
and with $\lambda (0) =$ 330 nm, obtained from the best fit to the anisotropic
order parameter described in the following paragraph. The curves are not too
far from each other, so there is no substantial difference for the choice
of  $\lambda (0)$ in this spatial range. Superfluid densities for both samples
A and B before and after electron irradiation are shown in \fref{fig2SI}.  Note
that both are normalized arbitrarily to 1 at $T=0$.   While it is clear that
electron irradiation results in a suppressed superfluid density at all temperatures,
we cannot make more rigorous conclusions, because $\lambda(0)$ definitely
increases with disorder, but at the moment we do not know how much.

To describe the data over the whole temperature range, we discuss fits using
a single anisotropic order parameter, as well as two isotropic gaps.  Neither
is really appropriate for a multiband, anisotropic superconductor, but these
analyses can give some sense of what properties the true  gap function must
display. In order to analyze the data with an anisotropic order parameter with
the possibility of both gapped and nodal states, we use a convenient
parameterization, $\Delta(t,\phi)=\Psi(t) \Omega(\phi)$, where the
temperature - dependent part, $\Psi(t)$, is obtained from the
self-consistency equation \cite{Kogan2009} and the angular
part, $\Omega (\phi)=(1+r \cos(4 \phi))/(1+r^2/2)^{1/2}$, is
chosen for a simple representation of the gap anisotropy.
Here $t=T/T_c$. In general, one could choose other anisotropic
harmonics, e.g. $\sim \cos(n \phi)$ with the symmetry of the
lattice\cite{ChubukovAR2012,Hirschfeld2016}, but this  would not
alter the qualitative results. The angular part is
normalized, $<\Omega^2>=1$. More details are given in \emph{Supplementary Information}.
A direct fit of the experimental $\rho_s(t)$ with $\lambda(0)=400$ nm to this
anisotropic gap can only reproduce the data roughly below 0.3$T/T_c$ with $r=$0.70.
However, a small adjustment of $\lambda(0)$ to 330 nm, produces a curve that can be
fitted with $r=$0.75 in the whole temperature range. The angular variation of the
gap is shown in the \emph{inset}(a) in \fref{fig5}. A hypothetical nodal case
with $r=1.2$ is shown for comparison. For the fitting, the temperature -
dependent part of the gap, $\Psi(t)$, was calculated self-consistently \cite{ProzorovKogan2011RPP}
and is shown in \emph{inset}(b) in comparison with the isotropic case of $r=0$.

 For completeness, we also used self-consistent two-gap $\gamma-$model \cite{gammaModel} as
 shown by the dashed line in \fref{fig5}, but being isotropic, it only captures
 the intermediate temperatures. Nevertheless, the interaction parameters inferred
 from the $\gamma-$model fitting result in a positive average of the
 interaction matrix, $<V> > 0$, which is important for the discussion below.

\section{Discussion: reconciling low-$T$ pairbreaking effects with $T_c$ enhancement by irradiation}
It is clearly important to try to reconcile data at low temperatures, including
the small gap and its enhancement with
electron irradiation -- all consistent with pairbreaking in an anisotropic  $s$-wave state --
with the remarkable  fact that $T_c$  {\it increases} with irradiation.
Note that there are several examples
in the literature where irradiation -- for example by heavy ions -- produces essentially
no change in $T_c$.  These effects have been understood in terms of mesoscopic inhomogeneity,
in contrast to the spatially uniform disorder produced
 at the nanoscale by electron irradiation.  In pnictides, however, e.g. the \ba-based ``Ba122" compounds, $T_c$ is suppressed
 fairly rapidly by 2.5 MeV electron irradiation \cite{Mizukami2014NatureComm,Prozorov2014PRX}.
 For example, in BaFe$_2$(As$_{0.67}$P$_{0.33}$)$_2$,  essentially the same irradiation dosage applied
 in this work produced a suppression of 5\% of $T_c$, whereas in FeSe $T_c$ increases by about the same amount.
 This effect is therefore qualitatively different.  Some aspects of the defects created
 by electron irradiation in this system are therefore not consistent with a purely
 pairbreaking interpretation, but may effectively dope the system, exert chemical pressure,
 or by some other means enhance the pairing interaction ("{\it pair strengthening}").
 Another possibility is if superconductivity in FeSe is competing with a secondary order
 that is suppressed more rapidly by disorder than superconductivity itself; this is
 analogous to the mechanism proposed for enhancement of $T_c$ by disorder in the
 spin density wave phase of the Ba(Fe$_{1-x}$Co$_x$)$_2$As$_2$ system \cite{Ni2010,Canfield2010review122}
 by Fernandes \etal\ \cite{TcEnhancement2012}.

Assuming that whatever effect leads to $T_c$ enhancement is rather small, one may ask,
how is it able to overcome the pairbreaking effect of disorder?   There are several
situations in which pairbreaking, even in a highly anisotropic superconductor, is fairly minimal.
The first example is a conventional non-sign-changing ``$s_{++}$ " superconductor, where
nonmagnetic disorder is pairbreaking only to the extent that it averages the gap
anisotropy \cite{MarkowitzKadanoff1963}. This seems unlikely simply because the electronic
interactions and Fermi surface of FeSe are so similar to the Fe-pnictides, where
there is considerable experimental  evidence and theoretical justification for
an $s_\pm$ identification\cite{Hirschfeld2016}.  The second is
an ``intraband" $s_\pm$ superconductor,  which behaves nearly
equivalently to ``$s_{++}$" in terms of non-magnetic scattering \cite{EfremovPRB2011}.  Here also, one would have to assume
attractive intraband interactions due, presumably, to phonons, leading to
a picture quite different from the other systems.  Finally, any sort
of $s_\pm$ pairing is fairly insensitive to disorder, at least as insensitive
as the corresponding anisotropic $s_{++}$ state,  provided the disorder
scattering is primarily intraband in nature.    It seems  to us that this latter
possibility is likely to be the case. If we compare to the example given above,
of electron irradiated BaFe$_2$(As$_{0.67}$P$_{0.33}$)$_2$, then the effect of
pair strengthening or competing order would have to be of roughly  the same order
but a bit larger compared to  the (opposite sign) effect of disorder pairbreaking.

Of the various scenarios considered to enhance $T_c$, some seem unlikely. For
example, we measured Hall coefficient in BaK122 crystals of different doping
levels and with different doses, and found that that electron irradiation is not doping
the system \cite{Prozorov2016TcTn}. Enhancement of $T_c$ by the suppression
of competing order by impurity scattering relies on a scenario whereby the competing
order is more sensitive to the disorder than the superconductivity itself. For example,
in the case discussed by Fernandes \etal\ \cite{TcEnhancement2012}, $(\pi,0)$ stripe order is
sensitive to impurity scattering by both $q=0$ and $(\pi,0)$, but isotropic $s_\pm$ superconductivity
is sensitive only to scattering by $(\pi,0)$.

In FeSe  there appears to be no long range magnetic order, but significant nematic order is present due to
weak orthorhombic distortion below structural transition.  Assuming a competition between the two states appears
reasonable  because $T_s$ is suppressed and $T_c$ enhanced both under hydrostatic pressure and,
more recently, sulfur doping \cite{MizuguchiTakano2009JPSJ_doping-effect,Mizuguchi2010,WatsonKim2015}.  The effect of disorder on these two competing states is however not as
straightforward as in the case of $s_\pm$ superconductivity competing with the $(\pi,0)$ spin density wave,
both because the nematic state  is a form of $q=0$ order, and because the anisotropic superconducting state is sensitive to small $q$ as well as large $q$ scattering. However, it can be shown that a $d-$wave Pomeranchuk instability is weakened by point-like impurity scattering \cite{Ho2008}. Such a suppression of a $d-$wave Pomeranchuk state is also expected in accordance with Imry-Ma theorem \cite{ImryMa1975}. Since $d-$wave Pomeranchuk order leads to deformation of the Fermi surface, it can strongly suppress superconductivity. If nematic order in FeSe is of this general type, we may expect that as it deteriorates due to disorder, superconductivity will get a boost, which under some circumstances may overcome the pairbreaking damage done by the disorder. Further work is needed to establish this scenario in context of multiband Fe-based superconductors.

For completeness, we mention that the nematic phase of FeSe has been interpreted in terms of various quadrupolar magnetic ``hidden" long range orders\cite{QSi2015,Nevidomskyy2016}, which may be quite  sensitive to disorder.  Thus far neither this sensitivity nor the competition with superconductivity has addressed in the literature.

We now consider the possibility that the Frenkel pairs created by
electron irradiation change the lattice in a way that mimics some kind
of chemical pressure, thereby altering the electronic structure and
thereby the pairing interaction itself subtly. One effect of this type
is of course actual hydrostatic pressure, where $T_c$ is observed to
increase simultaneously with the decrease of $T_s$, exactly as observed here.
On the other hand, the creation of Frenkel defects should expand rather
than collapse the lattice. Nevertheless, similar effects have been seen
when the lattice is expanded, e.g., in the FeSe intercalate family.  As
pointed out by Noji \etal\ \cite{Hosono2014}, expanding the lattice in
the $c-$ direction in the range of 5-9 \AA\ increases $T_c$ linearly at
a rate of about 14 K/\AA. FeSe itself is at the bottom of this
lattice constant range.  This trend in the intercalates was reproduced
by spin fluctuation theory with the calculated Fermi surfaces as
input\cite{Guterding2015}, and arises crudely due to the increase
of the Fermi level density of states as $c$ increases.  On the other hand,
uniaxial thermal-expansion measurements show, via thermodynamic relations, that \tc\ is mostly affected by the in-plane lattice parameters, $a$ and $b$, and is much less sensitive to the $c-$axis lattice constant \cite{Bohmer2013}. In either case, our estimates of the average stretch of the $c$-axis lattice lattice constant with irradiation provide an effect that is an order of magnitude too small to influence $T_c$ via chemical
pressure mechanism compared to the 5\% enhancement observed. With our irradiation dose, we create approximately 3.6$\times$10$^{-3}$ Frenkel pairs
per unit cell and even most optimistic estimates give a minuscule volume change, $\Delta V/V_0 < $10$^{-3}$, which at best can result in about 0.1 K change of $T_c$ for any optimistic scenario of either expansion of the $c-$axis \cite{Hosono2014} or hydrostatic pressure. Furthermore, upon warming up to room temperature about 30 \% of Frenkel pairs recombine as was directly determined from \emph{in-situ} resistivity measurements \cite{Prozorov2014PRX} and it is also believed that most interstitials will migrate from sample interior to surfaces, dislocations and other ``sinks" in the crystal \cite{Damask1963,THOMPSON1969}. This will make the above estimates even lower and we may safely conclude that pressure due to electron irradiation cannot explain our results.

This leaves us with the very plausible possibility that the impurity is pair strengthening, i.e. that it enhances the pair interaction locally, as discussed in several microscopic models \cite{Nunner2005,Maska2007,Foyevstova2009,Romer2012}.  Here the basic idea is that the electronic structure is modulated locally so that it enhances the magnetic exchange in the strong coupling limit, or drives the system closer to a local Stoner instability, in the weak coupling case. Note that the impurity can at the same time possess an electrostatic potential component that is itself pair breaking; the competition between these two effects decides whether $T_c$ is enhanced locally or not.  As discussed in \emph{Supplementary Information}, for the concentration of defects estimated in our irradiated sample, the defects are on the average well within a coherence length of one another, so there is a percolating superconducting path at the enhanced Tc, such that it can be detected in transport. The broadening of the transition by irradiation tends to support an inhomogeneous enhancement of this type. Note that since the above theoretical works considered only Hubbard-type 1-band models, considerable further work is necessary to establish the validity of this scenario in the context of the Fe-based materials.

\subsection{The role of twins}
The observation of small gaps at low temperatures in some samples may also be due to differences in sample growth, preparation and mounting for measurements that introduce different amounts of strain, and hence result in higher density of twins below the structural transition in some samples. Since twins appear to promote nodeless over nodal behavior\cite{Watashige2015} and the effect is long-range, samples with higher twin density may display predominantly nodeless gaps. It is interesting to note in this context  that the difference in resistivity between the nodal samples of Kasahara \etal\ and other samples (ours including) are mostly visible below the structural transition where twins form \cite{KnonerPRB2015}.

\section{Conclusions}
We have performed penetration depth measurements down to low temperatures
on pristine and electron-irradiated samples of FeSe.  In both samples, the
low-$T$ variation of $\Delta\lambda$ is consistent with a small minimum gap,
which increases from  0.7 K in the pristine sample to 1.3K in the irradiated sample,
suggesting the effect of gap averaging by disorder.  There are now reports in
the literature claiming both nodal and small full gaps in FeSe crystals, and it will
be important to establish whether the full gap samples are dirtier
or cleaner.  Thus far, our results with a single irradiation dose suggest that the gap opens with disorder, hence we expect the nodal samples are cleaner. Our findings of the small gap are
consistent with a highly anisotropic
gap function, either of $s_{++}$ or $s_{+/-}$ character, provided in the latter case disorder is of a sufficiently intraband character, so that \tc\ suppression is small.

At higher temperatures, we found that irradiation decreased the structural
transition $T_s$ by 0.9K, but surprisingly, $T_c$ was {\it enhanced} in the same
sample by 0.4K, nearly 5\% of $T_c$.   We discussed several theoretical scenarios that might
account for the latter effect, and concluded that a local pair strengthening
by irradiation-induced Frenkel defects, which locally enhance spin fluctuations near
a magnetic transition, is the most likely explanation.

\section*{Acknowledgements}
We thank A. Chubukov, B. Andersen, A. Golubov, R. Fernandes,
M. Iavarone, S. Maiti, Y. Matsuda, I. Mazin, S. Roessler,
T. Shibauchi, and L. Taillefer for useful discussions. We are particularly grateful to S. R\o{\ss}ler for drawing our attention to the effects of twins on the resistivity and to M.~Ko\'nczykowski as well as the whole SIRIUS team, B. Boizot, V. Metayer, and J. Losco, for running electron irradiation at \emph{\'{E}cole Polytechnique}.

This work was supported by the U.S. Department of Energy (DOE), Office of Science, Basic Energy Sciences, Materials Science and Engineering Division. Ames Laboratory is operated for the U.S. DOE by Iowa State University under contract DE-AC02-07CH11358. Electron irradiation was supported by EMIR network, proposal 11-11-0121. VM acknowledges the support from the Laboratory Directed Research and Development Program of Oak Ridge National Laboratory, managed by UT-Battelle, LLC, for the U. S. Department of Energy. PJH was partially supported by NSF-DMR-1005625. A. E. B. acknowledges support from the Helmholtz Association via PD-226.

\bibliographystyle{apsrev4-1}


%%%%%%%%%%%%%%%%%%%%%%%%%%%%%%%%%%%%%%%%%%%%%%
%%%%%%%%%%%%%%%%%%%%%%%%%%%%%%%%%%%%%%%%%%%%%%
\clearpage
\newpage

\chapter{Supplementary Information}
\setcounter{figure}{0}
\setcounter{equation}{0}

\makeatletter
\renewcommand{\thefigure}{S\@arabic\c@figure}
\makeatother

\makeatletter
\renewcommand{\theequation}{S\@arabic\c@equation}
\makeatother
\onecolumngrid

\section{Electron irradiation}

We measure the total dose of electron irradiation in \ccm\, by counting
the total charge that passed through a unit area. Therefore, 1 \ccm\
corresponds to $6.24 \times 10^{18}$ electrons per cm$^2$. In this work we accumulated 1.8 \ccm\ or $1.12 \times 10^{19}$ $e^-$/cm$^2$.

\begin{figure}[htb]
	\includegraphics[width=10cm]{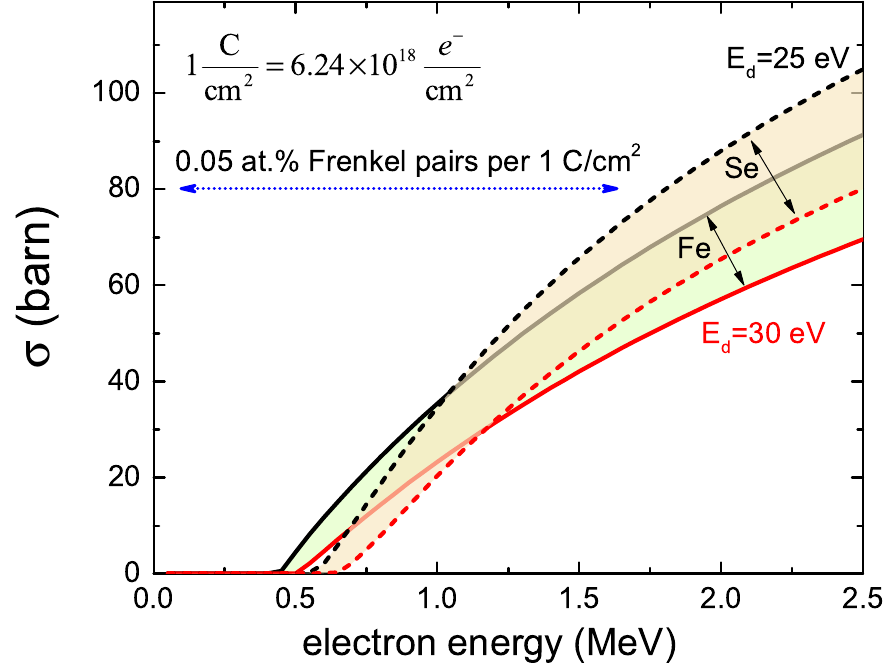}
	\caption{(Color online)  Ion - specific cross-sections calculated by using
	SECTE program. The upper curves correspond to the ion knock-out threshold
	of $E_d = $25 eV, the lower curves to $E_d = $30 eV. Se is shown by
	dashed curves, Fe by solid curves. At a mid-range value of 80 barn, we expect 0.05 at.\% per ion type of the Frenkel pairs per formula.}
	\label{fig1SI}
\end{figure}

\Fref{fig1SI} shows ion - specific cross - sections calculated by using the SECTE
simulation package \cite{SECTE}. Two pairs of curves for Fe (solid lines) and Se
(dashed lines) for two values of the displacement energy, $E_d = $25 eV (upper curves)
and 30 eV (lower curves), in the range commonly found in studies of various materials \cite{Damask1963,THOMPSON1969}.

At the energy of the electrons used in this study, 2.5 MeV, an average cross-section
of 80 barn will result in generation of 0.05 at.\% of Frenkel pairs for ions of each
kind per 1 \ccm of irradiation dose or 0.1 at.\% of defects of either kind per formula or, with Z=2 of FeSe, 0.2 at.\% of defects of either kind per unit cell.

It is instructive to compare the average distance between the
defects with the coherence length. A unit cell volume of FeSe
is 78.4 \AA$^3$. With Z=2, we have for 1.8 \ccm, 2$\times$1.8$\times$1$\times$10$^{-3}$=0.0036
Frenkel pairs of either Fe or Se per unit volume. Therefore, a volume
that will contain at least one Frenkel pair is 78.4/0.0036=2.1$\times$10$^4$ \AA$^3$,
so that the average distance between these defects is (2.1$\times$10$^4$)$^{1/3} \approx $=30 \AA. (Taking into account annealing of the defects upon warming up does not change this number much due to 1/3 power.)
This should be compared to the coherence length, which we can evaluate
from the upper critical fields. Along the $c-$axis, $H_{c2,c} \approx$ 17 T
and along the $ab-$plane it is about 30 T \cite{Terashima2014}. This gives
coherence lengths of $\xi_{ab}=110$ \AA\ and $\xi_{c}=83$ \AA, respectively.
Terashima \etal\ estimate the coherence lengths from the
slope of $dH_{c2}/dT$ at 130 \AA\ and 57 \AA, which is, indeed,
close to our estimate. In either case, these coherence lengths are
larger than the distance between the defects of 30 \AA\ and, therefore,
according to Markowitz and Kadanoff $T_c$ suppression should saturate as
function of scattering \cite{MarkowitzKadanoff1963}.

\section{Superfluid density of two samples before and after irradiation}

\begin{figure}[htb]
\includegraphics[width=10cm]{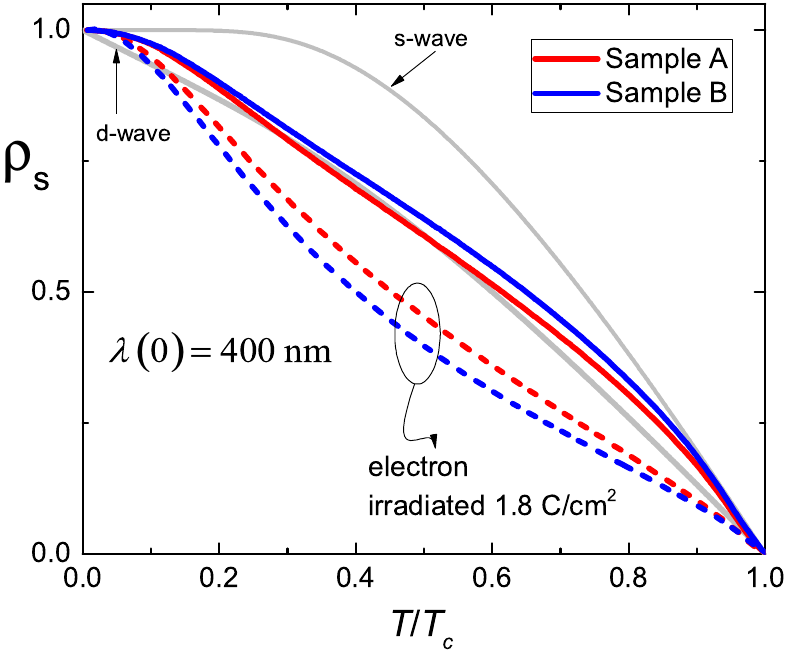}
\caption{(Color online) Superfluid density $\rho_s(t)$
before (solid lines) and after (dashed lines) 1.8 \ccm\ 2.5 MeV
electron irradiation for samples A and B. Solid grey lines show
standard $s-$ and $d-$ wave curves for comparison. $\lambda (0) =$ 400 nm was assumed.}
\label{fig2SI}
\end{figure}

In order to calculate superfluid density, $\rho_s=\left(1+\Delta \lambda (T)/\lambda(0)\right)^{-2}$,
from our data we need to know the absolute value of the London
penetration depth, $\lambda(0)$. In \fref{fig2SI}, the
superfluid density $\rho_ (t)$ is plotted with $\lambda (0) =$
400 nm obtained from microwave cavity perturbation measurements
of similar FeSe crystals \cite{Kasahara2014}. Saturation at $T \to 0$
can be clearly seen below approximately 0.5$T/T_c$ (as opposite to 0.3$T/T_c$
of the isotropic $s-$wave shown by grey line). An overall shape of $\rho_s(T)$
indicates a large number of thermally excited quasiparticles at the elevated
temperatures compared to the expectations of the isotropic gap and not too far
from the nodal $d-$wave line, but only at elevated temperatures. After electron
irradiation, the superfluid density decreases even more departing significantly
from the pristine samples. However, most likely $\lambda(0)$ increases, which will reduce the difference.

\section{Anisotropic $s-$wave gap with nodes}

We use a commonly - used ansatz of temperature and angle separation \cite{Kogan2009},

\begin{equation}\label{GapGeneral}
\Delta (T,\phi)=\Psi (T) \Omega (\phi),\quad \left< \Omega^2 \right>_{FS}=1
\end{equation}

Here we specifically use the form of the angular part of the gap commonly used to describe iron pnictides- based superconductors
\cite{ ChubukovAR2012,ChubukovHirschfeld2015,Hirschfeld2016},

\begin{equation}\label{Gap}
  \Omega (\phi) = \frac{1+r \cos4 \phi}{\sqrt{1+r^2/2}}
\end{equation}

\noindent and self-consistency equation for the temperature - dependent part, $\Psi(T)$, (see Eq.~(20) in Ref.~\cite{ProzorovKogan2011RPP}),

\begin{equation}\label{Psi}
  \frac{1}{2 \pi T}\ln{\frac{T_c}{T}}=\sum_{\omega>0}^{\infty}\left( \frac{1}{\hbar \omega} -\left<\frac{\Omega^2}{\sqrt{\Psi^2 \Omega^2+\hbar^2 \omega^2}} \right>_{FS} \right).
\end{equation}
\noindent where $\hbar \omega = \pi k_B T (2 n +1)$ are the Matsubara frequencies.
Fitting of the experimental superfluid density using Eqs.~(\ref{GapGeneral}), (\ref{Gap}) and (\ref{Psi}) is shown in \fref{fig5},
where \emph{inset}(a) shows the angular part, $\Omega(\phi)$ and \emph{inset}(b) shows self-consistent solutions for the gap, using Eq.~(\ref{Psi}).

\begin{figure}[htb]
	\includegraphics[width=9cm]{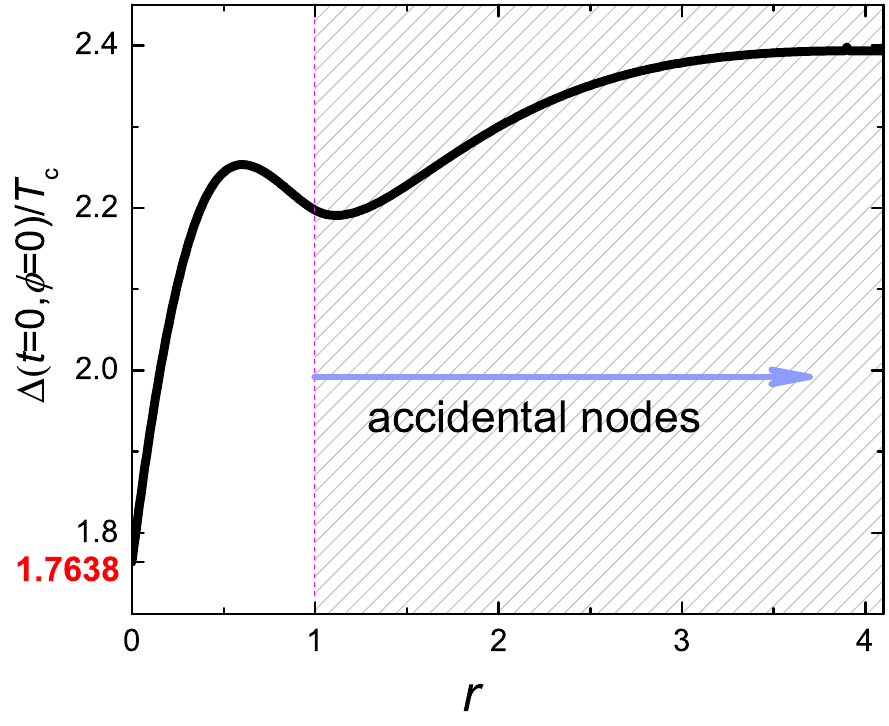}
	\caption{(Color online) $\Delta(t=0,r,\phi=0)/T_c$ vs $r$, where $\Delta(t,r,\phi)=\Psi(t)\Omega(r,\phi)$ and $\Psi(t)$
	is the solution of self - consistency equation, Eq.~\ref{Psi}.}
\label{fig3SI}
\end{figure}

\Fref{fig3SI} shows an interesting results of \emph{non-monotonic} $r-$ dependence
of the $\Delta(t=0,r,\phi=0)/T_c$, obtained self-consistently from Eq.~(\ref{Psi}). While
by itself it does not imply non-monotonic $T_c$, further microscopic analysis would be of interest.

Finally, the  Abrikosov-Gorkov
theory generalized to arbitrary $\Omega (\phi)$ (Born limit and isotropic scattering) reads \cite{Kogan2009},

\begin{equation}\label{AG}
-\ln{t_c}=\psi(\frac{1}{2}+\frac{g+g_m}{2 t_c})-\psi(\frac{1}{2})-<\Omega>^2 \left(\psi(\frac{1}{2}+\frac{g+g_m}{2 t_c})-\psi(\frac{1}{2}+\frac{g_m}{2 t_c}) \right)
\end{equation}
\noindent where $\psi$ is the digamma function, $t_c = T_c/T_{c0}$ and
normalized non-magnetic, $g=\hbar/(2 \pi k_B T_{c0} \tau )$, and
magnetic, $g_m=\hbar/(2 \pi k_B T_{c0} \tau_m )$ scattering rates
with $\tau$ and $\tau_m$ being potential and spin-flip scattering times,
respectively. In our case of $\Omega(\phi)$ given by Eq.~(\ref{Gap}), therefore $<\Omega>^2=2/(2+r^2)$ for a cylindrical Fermi surface.

\begin{figure}[htb]
	\includegraphics[width=8cm]{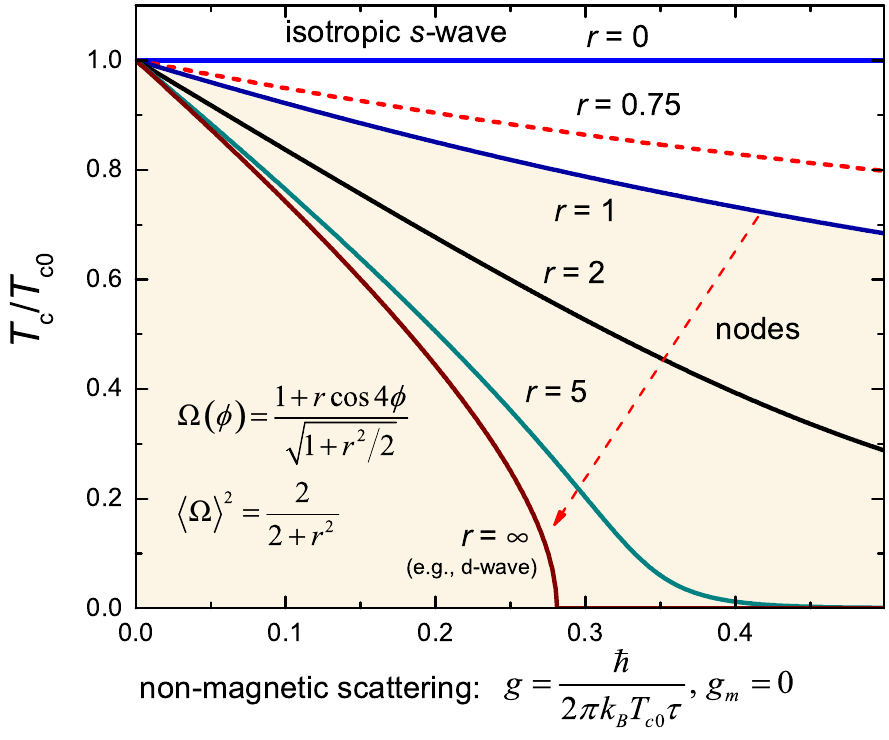}
	\caption{(Color online) Suppression of the transition
	temperature, $T_c/T_{c0}$, versus non-magnetic scattering rate, $g$,
	for several values of the anisotropy amplitude, $r$. At $r=$0 the gap
	is isotropic and accidental nodes appear for $r \geq$1. Best fit to our
	data, \fref{fig5}, is obtained with $r=$0.75 (red dashed line). $d-$wave state is shown for comparison.}
	\label{fig4SI}
\end{figure}

Solutions of Eq.(\ref{AG}) with $g_m$=0 and $\Omega$ given by Eq.~(\ref{Gap}) for different values of $r$ are shown in \fref{fig4SI}.

\end{document}